\documentclass[twocolumn,prb,superscriptaddress,floatfix,preprintnumbers,amssymb,amsmath]{revtex4-2}
\usepackage{graphicx}
\usepackage{dcolumn}
\usepackage{bm}
\usepackage[latin1]{inputenc}
\usepackage[mathscr]{eucal}
\usepackage{epsfig}

\usepackage{color,ulem}

\begin{document}

\title{Radio frequency Coulomb blockade thermometry}

\author{Florian Blanchet}
 \email{florian.blanchet@aalto.fi}
\author{Yu-Cheng Chang}
\author{Bayan Karimi}
\author{Joonas T. Peltonen}
\author{Jukka P. Pekola}
\email{jukka.pekola@aalto.fi}
\affiliation{Pico group, QTF Centre of Excellence, Department of Applied Physics,
    Aalto University, P.O. Box 15100, FI-00076 Aalto, Finland}

\date{\today}

\begin{abstract}

 We present a scheme and demonstrate measurements of a Coulomb blockade thermometer (CBT) in a microwave transmission setup. The sensor is embedded in an $LCR$ resonator, where $R$ is determined by the conductance of the junction array of the CBT. A transmission measurement yields a signal that is directly proportional to the conductance of the CBT,thus enabling the calibration-free operation of the thermometer. This is verified by measuring an identical sensor simultaneously in the usual dc setup. The important advantage of the rf measurement is its speed: the whole bias dependence of the CBT conductance can now be measured in a time of about 100\,ms, which is thousand times faster than in a standard dc measurement. The achieved noise equivalent temperature of this first rf primary measurement is about 1 mK/$\sqrt{{\rm Hz}}$ at the bath temperature $T=200\,$mK.
\end{abstract}

\maketitle

\section{Introduction}
Traditional single-electron tunneling phenomena are present when temperature $T$ is much lower than the unit of capacitive charging energy $E_{\rm C}$ determining the energy cost of adding or removing an electron in the device~\cite{Averin91,Ingold92}. In this Coulomb blockade regime, $k_{\rm B}T\ll E_{\rm C}$, charge transport is independent of temperature. Yet in the opposite regime, $k_{\rm B} T \gtrsim E_{\rm C}$, one obtains a thermometer with attractive properties~\cite{Pekola94,Pekola97,Bergsten,Nikolai-2,Hahtela20}. Coulomb blockade thermometer (CBT) is typically formed of an array of tunnel junctions, and it provides calibration-free (primary) thermometry in its operation range, temperature being deduced from a simple conductance measurement. Until today CBTs have been used as a reliable thermometer in the temperature range from $200\,\mu$K up to 60\,K~\cite{Sarsby20,Palma17,Bradley16,Meschke16,Jones20,Zumbuhl}, and in magnetic fields up to 27\,T~\cite{Pekola98}. To have a measure of absolute temperature, one needs to map the whole dependence of the differential conductance at bias voltage values ranging over several $k_{\rm B}T/e$. This is straightforward but in standard dc or quasi-dc measurements this takes typically several minutes in order to reach satisfactory accuracy in thermometry. Another option to measure conductance is to monitor either reflectance or conductance in a resonant circuit with the device-under-study embedded~\cite{Cleland,Karimi18,Prance,Imtiaz}. In this paper, we propose and demonstrate experimentally an rf transmission measurement of a CBT, where similar accuracy is achieved in less than one second. Transmitted microwave power is related to the conductance in a simple linear way, making the analysis and the determination of temperature from the measured signal straightforward. 
\section{Basics of the thermometer operation}
\begin{figure*}
	\centering
	\includegraphics [width=\textwidth] {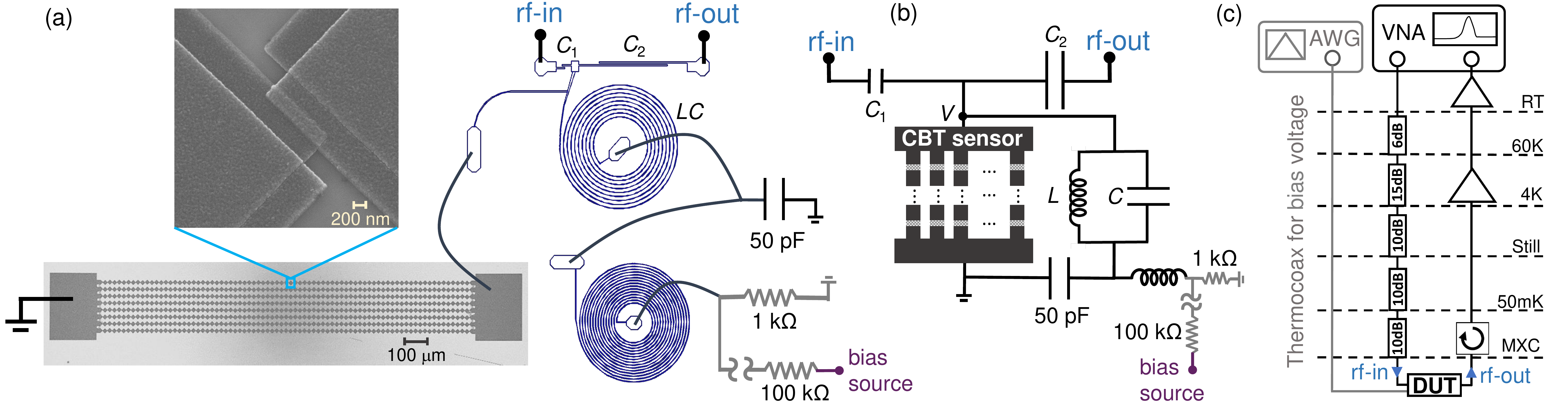}
	\caption{The setup for rf temperature measurement. (a) The device under study (DUT) together with the resonant circuit. The CBT sensor (scanning electron image on the left with a zoom-out of one tunnel junction) forms the dissipation $R$ of the spiral resonator (optical micrograph on the top right) in form $R=G(v)^{-1}$, where $G(v)$ is the bias dependent differential conductance of the junction array. $G(v)$ is obtained from the transmission $S_{21}$ between rf-in and rf-out ports, via coupling capacitors $C_1,C_2$, respectively. The dc bias across the CBT array is applied via an on-chip bias-tee formed of another superconducting spiral inductor on the bottom right. Physically the CBT is on one chip, and the rest of the components are on another one, apart from the $100\,{\rm k\Omega}$ bias resistor at room temperature, and the $1\,{\rm k\Omega}$ one and the $50\,$fF capacitor off the chip on the sample holder. The spiral resonators are made out of Nb. (b) Equivalent circuit of the panel (a). (c) The experimental rf setup for the CBT measurement outside the sample stage, including the low temperature and room temperature components. 
		\label{Fig1}}
\end{figure*}
When one measures the current $(I)$ and voltage across the array $(V)$ characteristics of a uniform series of $N$ tunnel junctions with normal metal electrodes, the asymptotic slope, i.e. conductance $G$ of this curve, at $|eV/N| \gg k_{\rm B}T,\,E_{\rm C}$, is given by $G_{\rm T}\equiv R_\Sigma^{-1}$, where $R_\Sigma$ is the total series resistance of the junctions and $V$ is the voltage across the whole array of junctions, see Fig.~\ref{Fig1}(a). At lower voltages, Coulomb repulsion of electrons on the islands lowers the conductance such that there is either full suppression of $G$ at $|eV/N| \lesssim E_{\rm C}$ at $k_{\rm B}T \ll E_{\rm C}$ or partial suppression at $|eV/N| \lesssim k_{\rm B}T$ for $k_{\rm B}T \gtrsim E_{\rm C}$, i.e. in the CBT operation range. Quantitatively, for $k_{\rm B}T \gtrsim E_{\rm C}$ the dependence of conductance on bias voltage can be cast in a universal form~\cite{Pekola97}
\begin{eqnarray} \label{e1}
G(v)/G_{\rm T}=&&1-u\,g(v)\\&&-\frac{u^2}{4}[g'(v)h'(v)+g''(v)h(v)],\nonumber
\end{eqnarray}
where $g(v)=e^v[e^v(v-2)+v+2]/(e^v-1)^3$, $u=E_{\rm C}/(k_{\rm B}T)$, $v= eV/(Nk_{\rm B}T)$, and $h(v)=v \coth(v/2)$. The charging energy $E_{\rm C}$ is given by the full capacitance matrix of the thermometer; for an array with junction capacitance $C$ for each of them and vanishing stray capacitance, $E_{\rm C}=(1-1/N)e^2/C$. The important property of Eq.~\eqref{e1} is that measuring the voltage half-width at full minimum, $V_{1/2}$, gives the absolute temperature without need to know the specific parameters of the sensor apart from the number of junctions in the array. The commonly adopted result valid for $u\ll1$ (the first line of Eq.~\eqref{e1}) tells that $V_{1/2}=5.4392Nk_{\rm B}T/e$, i.e. 
\begin{equation} \label{e2}
T=0.18385 \frac{e}{Nk_{\rm B}}V_{1/2}.
\end{equation}
The numerical prefactor arises from the functional form of $g(x)$. The depth of the zero-bias dip, $\Delta G \equiv G_{\rm T}-G(0)$, can also serve as a thermometer, since $\Delta G/G_{\rm T}\approx u/6$, (again up to linear order in $u$) but here the calibration depends on the capacitances via $E_{\rm C}$ (secondary thermometer). The corrections to Eq.~\eqref{e2} arising from the second line in Eq.~\eqref{e1} lead to~\cite{Pekola97} 
\begin{equation} \label{e3}
	T=0.18385 \frac{e}{Nk_{\rm B}}V_{1/2}/(1+0.3921 \Delta G/G_{\rm T}),
\end{equation}
where the correction, $(1+0.3921 \Delta G/G_{\rm T})^{-1}$, can be easily determined from the measured depth $\Delta G/G_{\rm T}$. This correction is about 2\% in $T$ at the lowest temperatures of the current measurement, and smaller above it. Equations~\eqref{e1}-\eqref{e3} are naturally valid only for a uniform array; corrections due to non-uniformities are discussed, e.g., in~\cite{Pekola97}. 

\section{rf measurement}
The actual sample and the measurement setup are presented schematically in Figs.~\ref{Fig1}(a) and~\ref{Fig1}(b), together with the full wiring, amplifiers and attenuators in~\ref{Fig1}(c). The heart of the system is an $LCR$ tank circuit that resonates at $f_0\approx 547\,$MHz. The CBT thermometer forms the parallel dissipation $R$ of this resonator. The $LC$ is formed of a superconducting (Nb) spiral. With the actual component values, $L=110\,$nH, $C=540\,$fF, and $R\approx R_\Sigma=80\,\rm k\Omega$ as the resistance of 80 junctions in series in 8 parallel chains (Sample CBT80, shown in Fig.~\ref{Fig1}(a)) to be described below. The quality factor of the resonator is $Q\approx 45$. This yields an internal response time of the measurement as $Q/(2\pi f_0)\sim 13$ ns which is far shorter than the typical data acquisition times in this work. The measured signal is the transmission from rf-in to rf-out (see Figs.~\ref{Fig1}(a) and~\ref{Fig1}(b)), through the coupling capacitances $C_1 \approx 67\,$fF and $C_2\approx 170\,$fF at the input and output, respectively. These values are obtained from Sonnet simulation and are also consistent with the mutual capacitance matrix simulated by Comsol. This transmission is affected by the differential conductance of the CBT sensor, which is placed physically on a different chip next to the one with the resonator. The dc bias voltage $V$ is applied via the bias-tee (shown in Figs.~\ref{Fig1}(a) and~\ref{Fig1}(b)  with grey colour), where the inductor is formed by another spiral filter made of Nb on the same chip as the main resonator.   

The transmission in the circuit of Fig.~\ref{Fig1}(b) (in dB) from rf-in to rf-out $S_{21}$ is given by~\cite{Karimi18}
\begin{equation} \label{e4}
S_{21}=S_0-20 \lg(1+R_0G).
\end{equation}
Here $S_0$ is a constant and $R_0\approx(R_{\rm L}(2\pi f_0C_2)^2)^{-1}\approx 59\,\rm k\Omega$, where $R_{\rm L}\approx 50\,\Omega$ is the line impedance. When $\frac{R_0G_{\rm T}}{1+R_0G_{\rm T}}\frac{\Delta G}{G_{\rm T}}\ll 1$, we may linearize Eq.~\eqref{e4} into
\begin{equation} \label{e6}
	S_{21}=\tilde S_0-\frac{20}{\ln(10)}\frac{R_0G_{\rm T}}{1+R_0G_{\rm T}}\frac{G(V)}{G_{\rm T}},
\end{equation}
where $\tilde S_0$ is another constant. For the current resonator and sensor $R_0G_{\rm T}\approx\,$0.7, i.e. the linear approximation is valid for all temperatures here, i.e. for $\Delta G/G_{\rm T}\ll 1$. Secondary temperature measurements at zero bias, i.e. measurements of $\Delta G /G_{\rm T}$ without a bias sweep, are certainly possible as well, since the grounding of the sensor at the base temperature gives extra stability to the biasing~\cite{Karimi18}. Due to the focus on primary thermometry, we do not present data on secondary measurements. 

\section{Description of the experiment}
In this work, we present a simultaneous measurement of three different CBT sensors in a dilution refrigerator in the temperature range around 200\,mK. One of the sensors, coined Sample CBT100, is a reference consisting of chains of 100 junctions in each with 10 parallel arrays. This CBT is made of Al and Cu with superconductivity suppressed by the field of a permanent magnet on it. It has large volume inter-junction islands to ensure proper thermalization. It has proven to provide accurate temperature in numerous earlier experiments in the range 10-500\,mK~\cite{Hahtela20}. This sensor is measured using standard lock-in techniques to obtain the differential conductance at low frequency ($\sim 30\,$Hz). 

\begin{figure}
	\centering
	\includegraphics [width=\columnwidth] {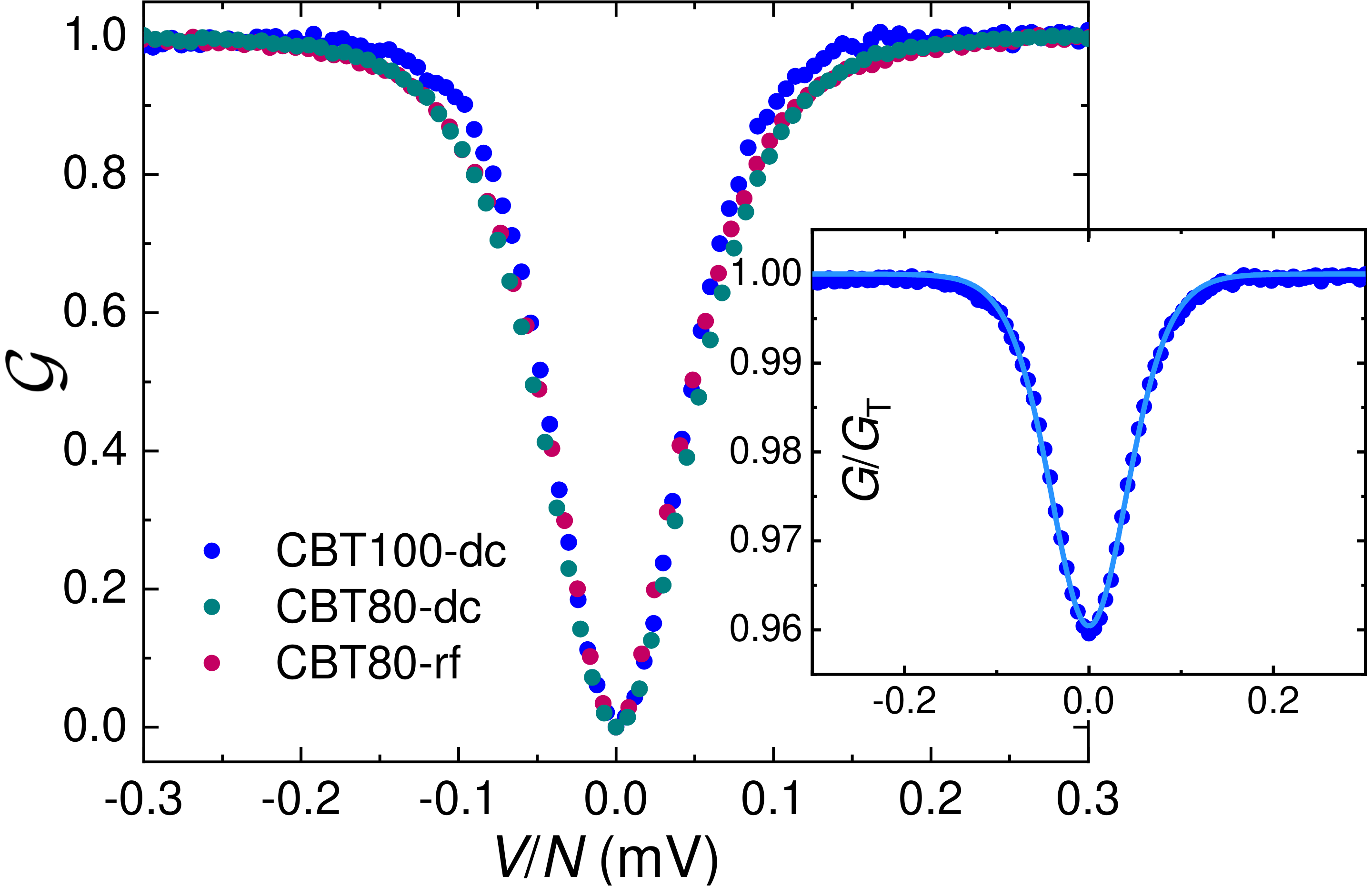}
	\caption{Results of the conductance measurements of the three sensors at $T\sim 208$\,mK versus voltage per junction, $V/N$. For convenient comparison, the conductance has been normalized to the range $0-1$ by presenting $\mathcal{G}\equiv[G(V)-G(0)]/[G_{\rm T}-G(0)]$. All the dips have approximately the same width and shape. The reference sensor (CBT100) data is shown by symbols as $G(V)/G_{\rm T}$ separately in the inset with a fit (line) from Eq.~\eqref{e1}, yielding $u=E_{\rm C}/(k_{\rm B}T)\sim 0.24$ for this measurement.
		\label{Fig2}}
\end{figure}
The other two CBT sensors (CBT80) are made of AlMn alloy to avoid superconductivity in the absence of magnetic field down to the lowest temperatures~\cite{Ruggiero04}. The Sample CBT80-dc is connected again to a low frequency lock-in circuit to measure the differential conductance. This serves as another reference sample for the main device of the current work, Sample CBT80-rf, which is the sensor connected to the rf-circuitry described above and shown in Fig.~\ref{Fig1}(a). These two latter sensors are made in the same fabrication batch with nominally identical characteristics. The CBT80-dc and CBT80-rf devices were fabricated on an oxidized Si substrate by two-angle shadow evaporation through a suspended, germanium hard mask~\cite{Pekola13} that was patterned by electron beam lithography. The devices were deposited in an electron beam evaporator using an alloy target with $0.3\,\%$ nominal Mn concentration. First, an approximately 20-nm-thick film was evaporated (30\,nm deposited at $-45^\circ$ tilt angle) at a rate of 3\,${\rm \AA/s}$, followed by in situ oxidation in 10\,mbar pure oxygen for 10 minutes. The tunnel junctions were completed by depositing the approximately 30-nm-thick second electrode (40\,nm at $+45^\circ$ tilt angle). The overlap area of each junction is approximately $1.2\,\mu{\rm m}\times 180\,{\rm nm}$. The lateral size of the island between neighboring junctions is $15\,\mu{\rm m}\times 15\,\mu{\rm m}$. This relatively small island size and the weak electron-phonon coupling in the AlMn alloys~\cite{Taskinen06a,Taskinen06b} limit the electron thermalization at temperatures below 200\,mK; hence we present experiments only above this range.

The CBT80-rf connects in parallel to a superconducting spiral resonator. The resonator is made of sputtered 200\,nm Nb film on a 675\,$\mu$m-thick high resistance intrinsic silicon wafer and patterned by electron-beam lithography and reactive ion etching process. For reading out the transmittance, the resonator is capacitively coupled to the input (output) microwave feed-lines as shown in Figs.~\ref{Fig1}(a) and~\ref{Fig1}(b), and could be easily integrated with an on-chip bias-tee and a divider on the sample holder. The sample is voltage biased by a triangular waveform at 5\,Hz from an arbitrary waveform generator (AWG) at room temperature through a thermocoax line to a 1\,k$\Omega$ surface-mount load resistor grounded on the sample holder. It is also connected in parallel with the CBT80-rf via the spiral coil of the on-chip bias-tee. The incident microwave signal comes via several attenuators distributed at varying temperatures in order to minimize the thermal noise from high temperatures before exciting the resonator. The coupled output voltage across the resonator is amplified by the high electron mobility transistor amplifier at 4\,K and a circulator with a bandwidth between 480\,MHz and 720\,MHz which isolates the sample from the amplifier noise by 20\,dB. The signal is amplified again at room temperature to ensure that it well exceeds the noise floor of the vector network analyzer (VNA). 

The change of the conductance of the CBT80-rf by the bias voltage is then reflected in the height and width of the transmittance peak. The fast measurement is done by biasing the CBT80-rf with triangular waveform with peak to peak voltage of 60\,mV using AWG and monitoring the transmittance at resonant frequency $f_0$ by VNA with intermediate frequency bandwidth of 1 kHz. The corresponding sampling time interval is 1.08\,ms. The data acquisition is synchronized with the triangular waveform through a transistor-transistor logic trigger from AWG to VNA with rising trigger repeating each 10 s.
\section{Results}
Figure~\ref{Fig2} demonstrates the key feature of the measurement, namely that there is no significant difference between CBT80-dc and CBT80-rf curves: the low frequency and the rf measurements then yield essentially the same temperature. The shoulder shapes are slightly different in the reference (CBT100-dc) and CBT80 samples. This is because of the non-optimized sensor design of the AlMn samples (CBT80-dc and CBT80-rf), i.e. the thermalization is limited by the smallness of the islands in between the junctions.

\begin{figure}
	\centering
	\includegraphics [width=\columnwidth] {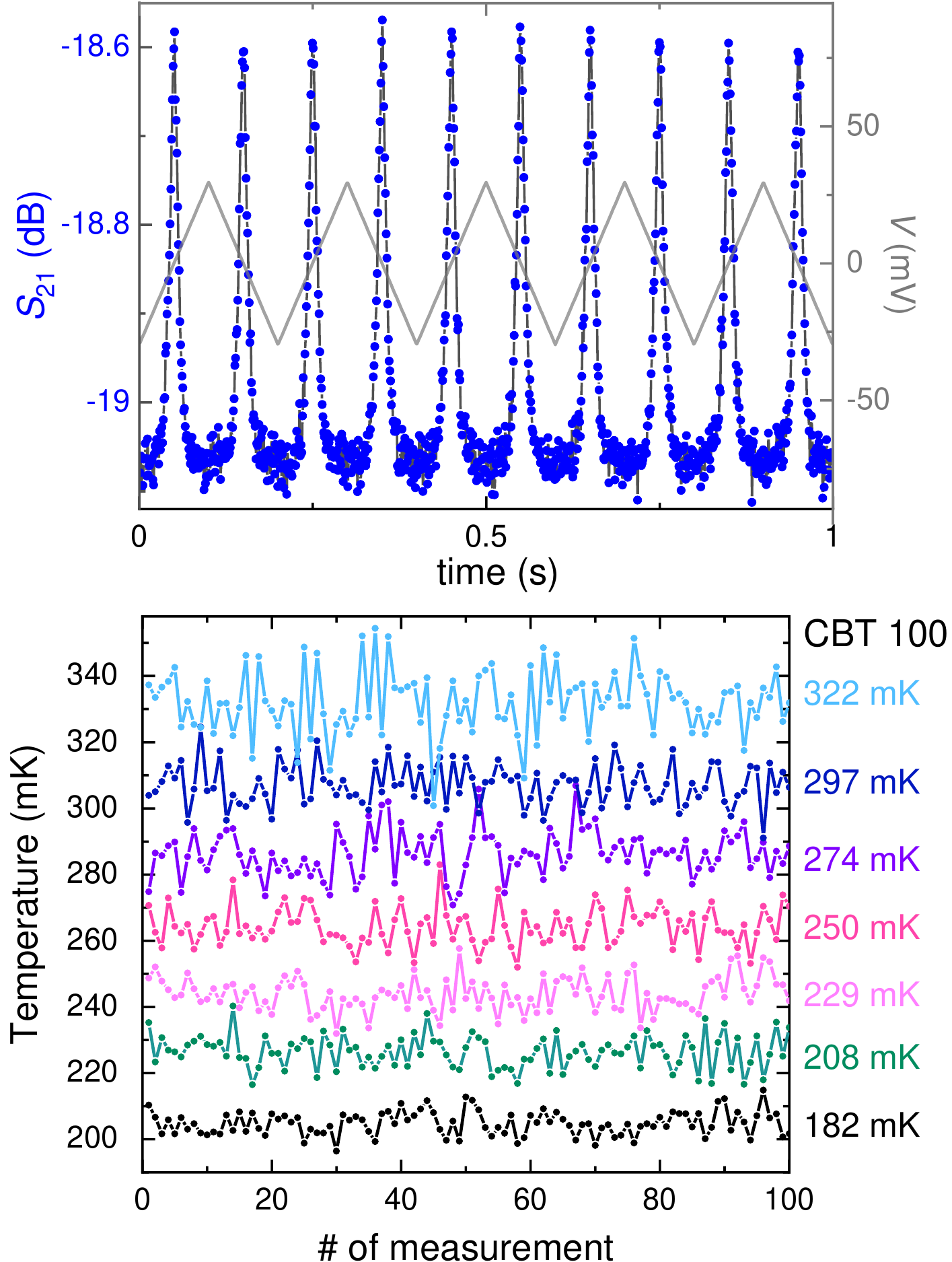}
	\caption{Transmission $S_{21}$ measurements. (Top) $S_{21}$ using a symmetric triangular voltage ramp in the range $\pm 30$\,mV across the array, repeating a half-period ramp each 100\,ms, measured at $T=182\,$mK. Successive peaks then correspond to opposite ramp directions. $S_{21}$ inverts the conductance dip, see Eq.~\eqref{e6}. (Bottom) $T$ determined at each measurement temperature (reference temperatures from CBT100-dc are indicated for each set of data on the right) from 100 peaks of the type in the top panel repeated once in 100 ms. From these sets we deduce noise equivalent temperature $\rm NET$ (see text) in the absolute (full curve) measurement ranging from ${\rm NET}=1...3$\,mK/$\sqrt{{\rm Hz}}$ from the lowest (200\,mK) to the highest (340\,mK) temperatures.    
		\label{Fig3}}
\end{figure}

The top panel of Fig.~\ref{Fig3} shows 10 successive transmission curves taken in 1\,s with the said triangular wave pattern as indicated by the light gray line in the figure. The widths of these peaks in $S_{21}$ then serve as the thermometer based on Eqs.~\eqref{e1} and \eqref{e6}. The temperature is obtained by finding the best fit of the form given by these equations for each half period of the triangular drive and determining the full-width at half-maximum for each curve. The bottom panel in Fig.~\ref{Fig3} shows the temperatures extracted from such fits for 100 successive peaks with alternative polarity of the linear sweep of $V$. At each bath temperature, indicated by the reference CBT100-dc thermometer reading on the right-hand-side, the sequence of temperatures indicates some scatter in this fast measurement. We achieve noise equivalent temperature NET of $\sim 1\,$mK/$\sqrt{\rm Hz}$ (a 1-sigma uncertainty of 2\,\% at the bath temperature of 200\,mK in 100\,ms). The value of NET is obtained as the square-root of the quantity found by multiplying the variance of $T$ for the 100 points in Fig.~\ref{Fig3} at each temperature by the measurement time (100\,ms here). The operating time of 100 ms is a thousand-fold improvement as compared to standard methods with measurement times in minutes in lock-in or direct dc detection.
\section{Discussion}
In this work, we have demonstrated an efficient way to measure sub-kelvin absolute temperatures by a Coulomb blockade thermometer coupled to a superconducting resonator. We envision some future improvements to the method, both from the point of view of low-temperature circuitry and room temperature instrumentation. 

In order to increase the signal-to-noise ratio, some sources of noise can be suppressed by standard means. First, in our device the coupling with the resonator is made with simple capacitors forming first order high-pass filters; the noise can be reduced further using a more complex circuitry for selective coupling, reducing the bandwidth or increasing the order. Apart from the sample, the noise can also be controlled by the choice of the fridge circuitry; in our experiment, a single large bandwidth circulator is used between the sample and the 4 K HEMT amplifier. Its purpose is to protect the sample from the backaction noise of the amplifier; it can be improved with a more selective circulator and/or by adding a second one with additional 20\,dB rejection. Ultimately, the use of a low-temperature amplifier, like Josephson parametric amplifier, should significantly reduce this noise; however it would also increase the complexity of the circuit.

Most of the room-temperature electronics can be packed in a complex single chip for easy-to-use commercial purposes. Here we use general purpose instruments but only few features of them are really used. First, a clock is needed for synchronization, both a low frequency triangular and high frequency sinusoidal generators rely on it. Moreover, working with telecommunication frequency, specific purpose integrated circuits exist to perform demodulation, and finally a regular FPGA can be programmed to extract the temperature as described in the text. We presented here a proof-of-concept technique for Coulomb blockade thermometry which will hopefully find applications both in basic science as well as in practical instruments.

\section{Acknowledgments}
This work was supported by Academy of Finland grant 312057 (QTF Centre of Excellence) and from the EMPIR programme co-financed by the Participating States and from the European Union's Horizon 2020 research and innovation programme. We acknowledge the provision of facilities and technical support by Aalto University at OtaNano - Micronova Nanofabrication Center and LTL infrastructure which is part of European Microkelvin Platform (EMP, No. 824109 EU Horizon 2020).

\end{document}